\documentclass[12pt,reqno]{article}
\usepackage{amsfonts}
\usepackage{amsmath}
\usepackage{amsbsy}

\usepackage{amssymb,latexsym}
\numberwithin{equation}{section}

\begin{document}
 \allowdisplaybreaks[1]
\title{Symmetric Space $\sigma$-model Dynamics: Internal Metric Formalism}
\author{Nejat T. Y$\i$lmaz\\
Department of Mathematics
and Computer Science,\\
\c{C}ankaya University,\\
\"{O}\u{g}retmenler Cad. No:14,\quad  06530,\\
 Balgat, Ankara, Turkey.\\
          \texttt{ntyilmaz@cankaya.edu.tr}}
\maketitle
\begin{abstract}
For the symmetric space sigma model in the internal metric
formalism we explicitly construct the lagrangian in terms of the
axions and the dilatons of the solvable Lie algebra gauge and then
we exactly derive the axion-dilaton field equations.
\end{abstract}

\section{Introduction}
It is well known that the sigma model lagrangian of the Riemannian
globally symmetric space $G/K$ can be formulated by using a
definition of an internal metric. The construction can be referred
in \cite{nej1,nej2,ker1,ker2,julia1,julia2}. The choice of the
solvable Lie algebra gauge \cite{fre} for parameterizing the coset
representatives brings further simplicity in the construction.
Under a specified trace condition the field equations of the
symmetric space sigma model are derived and further studied in
\cite{ker1,ker2} and \cite{nej1,nej2} respectively. However since
the lagrangian is not explicitly constructed in terms of the coset
scalars the formulation of \cite{ker1,ker2} is based on the
Lagrange multiplier methods and the field equations are written in
terms of the field strengths of the axions which are treated as
independent fields. It is also mentioned in \cite{ker1,ker2} that
if one can express the lagrangian in terms of the scalar fields
explicitly one can directly vary it to obtain the exact field
equations of the dilatons and the axions. On the other hand in
\cite{nej1,nej2} the Cartan-forms in terms of which the symmetric
space sigma model lagrangian can be expressed are calculated
exactly. This promises an explicit formulation of the symmetric
space sigma model lagrangian in terms of the coset scalar fields
and the derivation of the field equations for a generic trace
convention.

In this work we go in the above mentioned direction to obtain the
most general form of the field equations of the sigma model which
is based on the Riemannian globally symmetric space $G/K$. By
using the exact form of the Cartan-form we express the SSSM
lagrangian explicitly in terms of the coset scalars in a generic
trace convention then by varying it directly we obtain the field
equations of the theory. We will assume the solvable Lie algebra
gauge to parameterize the coset manifold $G/K$ and we will
classify the scalar fields as axions and the dilatons referring to
the non-perturbative effective string theory and the supergravity
literature where the symmetric space sigma model plays a central
role governing the scalar sector which reveals the global symmetry
and the U-duality structure of the supergravity and string
theories respectively \cite{nej125,nej126}.

In section two leaving some of the details to the references we
will present a concise formulation of the lagrangian for the
axion-dilaton parametrization. Without choosing a specific trace
convention which generalizes the formalism of
\cite{nej1,nej2,ker1,ker2} we will construct the lagrangian
explicitly in terms of the scalar fields for a generic trace
convention. Then in section three we will vary the symmetric space
sigma model lagrangian to derive the general field equations of
the axion and the dilaton scalar fields.
\section{Lagrangian in the Axion-Dilaton
Parameterization}\label{section1} The construction of the
symmetric space sigma model is based on a set of $G$-valued maps
$\{\nu(x)\}$ which are onto $C^{\infty}$-maps from the
$D$-dimensional spacetime to the coset space $G/K$. Thus they
parameterize the coset manifold $G/K$. Here $G$ is in general a
non-compact real form of any other semi-simple Lie group and $K$
is a maximal compact subgroup of $G$. The coset manifold $G/K$ has
a unique analytical structure induced by the quotient topology of
$G$. The scalar manifold $G/K$ is a Riemannian globally symmetric
space for all the $G$-invariant Riemannian structures on $G/K$
\cite{hel}. The solvable Lie algebra gauge is a parametrization of
the coset manifold $G/K$ which is due to the Iwasawa decomposition
\begin{equation}\label{s1}
\begin{aligned}
g&=\mathbf{k}\oplus \mathbf{s}\\
\\
&=\mathbf{k}\oplus \mathbf{h_{p}}\oplus \mathbf{n},
\end{aligned}
\end{equation}
of the Lie algebra $g$ of $G$. In \eqref{s1} $\mathbf{k}$ is the
Lie algebra of $K$ and $\mathbf{s}=\mathbf{h}_{p}\oplus
\mathbf{n}$ is a solvable Lie subalgebra of $g$. The abelian
subalgebra $\mathbf{h}_{p}$ is generated by $r$ non-compact Cartan
generators $\{H_{i}\}$. Also the nilpotent Lie subalgebra
$\mathbf{n}$ is generated by a subset $\{E_{m}\}$ of the positive
root generators of $g$ where $m\in\Delta_{nc}^{+}$. The roots in
$\Delta_{nc}^{+}$ are the non-compact roots with respect to the
Cartan involution associated with the Iwasawa decomposition
\eqref{s1} \cite{nej1,nej2,ker1,hel}. The map
\begin{equation}\label{s2}
Exp: \mathbf{s} \longrightarrow G/K,
\end{equation}
from the ${\Bbb{R}}^{dim\mathbf{s}}$-manifold $\mathbf{s}$ into
$G/K$ is a local diffeomorphism \cite{hel}. Therefore
\begin{equation}\label{s3}
\nu (x)=e^{\frac{1}{2}\phi ^{i}(x)H_{i}}e^{\chi ^{m}(x)E_{m}},
\end{equation}
is a legitimate parametrization of the coset manifold $G/K$ which
is called the solvable Lie algebra gauge \cite{fre}. The scalar
fields $\{\phi^{i}\}$ are called the dilatons and $\{\chi^{m}\}$
are called the axions. In the internal metric formalism
\cite{nej1,nej2,ker1,ker2,julia1,julia2} of the symmetric space
sigma model the lagrangian which is invariant under the right
rigid action of $G$ and the left local action of $K$ is
constructed as
\begin{equation}\label{s4}
 \mathcal{L}=\frac{1}{4}\, tr(\ast d{\mathcal{M}}^{-1}\wedge
 d{\mathcal{M}}),
\end{equation}
where the internal metric $\mathcal{M}$ is defined as
\begin{equation}\label{s5}
 \mathcal{M}=\nu ^{\#}\nu.
\end{equation}
The generalized transpose $\#$ over the Lie group $G$ is such that
$(exp(g))^{\#}=exp(g^{\#})$. It is defined by using the Cartan
involution $\theta$ over $g$ that is associated with \eqref{s1} as
$g^{\#}=-\theta(g)$ \cite{hel}. If the subgroup of $G$ generated
by the compact generators is an orthogonal group then in the
fundamental representation of $g$ we have $g^{\#}=g^{T}$. Also it
is always possible to find a matrix representation of $g$ in which
$\#$ coincides with the matrix transpose operator \cite{ker1}. In
spite of the fact that the generalized transpose $\#$ shares the
usual properties of the matrix transpose in our formulation we
will assume a matrix representation in which $g^{\#}=g^{T}$. From
the definition of the coset parametrization in \eqref{s3} we have
the identities
\begin{equation}\label{s6}
\nu ^{-1}d\nu=-d\nu ^{-1}\nu\quad ,\quad d\nu\nu ^{-1}=-\nu d\nu
^{-1}.
\end{equation}
Also
\begin{equation}\label{s7}
tr(d\nu_{1}\wedge \ast d\nu_{2})=(-1)^{(D-1)}tr(\ast
d\nu_{2}\wedge d\nu_{1}),
\end{equation}
for two matrix-valued functions $\nu_{1}$ and $\nu_{2}$. Now if we
define the Cartan-Maurer form $\mathcal{G}$ as
\begin{equation}\label{s8}
{\mathcal{G}}=d\nu \nu ^{-1},
\end{equation}
in the light of the above mentioned identities, the properties of
the coset representatives and the generalized transpose we can
express the lagrangian \eqref{s4} in terms of the Cartan-Maurer
form $\mathcal{G}$ as
\begin{equation}\label{s9}
{\mathcal{L}}=-\frac{1}{2}\, tr(\ast \mathcal{G}\wedge
\mathcal{G}^{\#}+\ast \mathcal{G}\wedge \mathcal{G}).
\end{equation}
The Cartan-Maurer form $\mathcal{G}$ is explicitly calculated in
terms of the axions and the dilatons in \cite{nej1,nej2}
\begin{equation}\label{s10}
\mathcal{G}=\frac{1}{2}d\phi ^{i}H_{i}+\overset{\rightharpoonup }{
\mathbf{E}^{\prime }}\:\mathbf{\Omega }\:\overset{\rightharpoonup
}{\mathbf{d\chi}}.
\end{equation}
The row vector $\overset{\rightharpoonup }{ \mathbf{E}^{\prime }}$
is
\begin{equation}\label{s11}
(\overset{\rightharpoonup }{ \mathbf{E}^{\prime }})_{\alpha}=e^{
\frac{1}{2}\alpha _{i}\phi ^{i}}E_{\alpha}.
\end{equation}
Also $\overset{\rightharpoonup }{\mathbf{d\chi}}$ is a column
vector of the field strengths of the axions $\{d\chi^{i}\}$. In
\eqref{s10} $\mathbf{\Omega}$ is a
dim$\mathbf{n}\times$dim$\mathbf{n}$ matrix
\begin{equation}\label{s12}
 \mathbf{\Omega}=(e^{\omega}-I)\,\omega^{-1}.
\end{equation}
The dim$\mathbf{n}\times$dim$\mathbf{n}$ matrix $\omega$ is also
defined as
\begin{equation}\label{s13}
 \omega _{\beta }^{\gamma }=\chi ^{\alpha }\,K_{\alpha \beta
}^{\gamma}.
\end{equation}
The structure constants $K_{\alpha \beta }^{\gamma }$ and the root
vector components $\alpha_{i}$ are defined as
\begin{equation}\label{s14}
[E_{\alpha },E_{\beta }]=K_{\alpha \beta }^{\gamma }\,E_{\gamma
}\quad ,\quad [H_{i},E_{\alpha}]=\alpha_{i}\,E_{\alpha }.
\end{equation}
Since now we have the exact form of the Cartan-Maurer form
$\mathcal{G}$ we can express the lagrangian \eqref{s4} explicitly
in terms of the axions and the dilatons. Inserting \eqref{s10} in
\eqref{s9} we obtain
\begin{subequations}\label{s15}
\begin{align}
{\mathcal{L}}&=-\frac{1}{8}\,\mathcal{A}_{ij}\ast
d{\phi}^{i}\wedge
 d\phi^{j}-\frac{1}{4}\,\mathcal{B}_{i\alpha}\ast d{\phi}^{i}\wedge
e^{\frac{1}{2}\alpha_{i}\phi^{i}}
U^{\alpha}\notag\\
\notag\\
&\quad
-\frac{1}{2}\,\mathcal{C}_{\alpha\beta}e^{\frac{1}{2}\alpha_{i}\phi^{i}}\ast
U^{\alpha}\wedge e^{\frac{1}{2}\beta_{i}\phi^{i}}
U^{\beta},\tag{\ref{s15}}
\end{align}
\end{subequations}
in which we slightly change the notation introduced above and use
\begin{equation}\label{s16}
U^{\alpha}=\mathbf{\Omega}^{\alpha}_{\beta}d\chi^{\beta}.
\end{equation}
For the sake of generality in \eqref{s15} we have not specified
any trace convention and we have defined the generic trace
coefficients as
\begin{subequations}
\begin{gather}\label{s17}
\mathcal{A}_{ij}=tr(H_{i}H_{j}^{\#})+tr(H_{i}H_{j}),\notag\\
\notag\\
\mathcal{B}_{i\alpha}=tr(H_{i}E_{\alpha}^{\#})+tr(E_{\alpha}H_{i}^{\#})+
tr(H_{i}E_{\alpha})+tr(E_{\alpha}H_{i}),\notag\\
\notag\\
\mathcal{C}_{\alpha\beta}=tr(E_{\alpha}E_{\beta}^{\#})+tr(E_{\alpha}E_{\beta})\tag{\ref{s17}}.
\end{gather}
\end{subequations}
By using the properties of the generalized transpose $\#$
$\mathcal{B}_{i\alpha}$ can further be expressed as
\begin{equation}\label{s17.5}
\mathcal{B}_{i\alpha}=2(tr(E_{\alpha}H_{i}^{\#})+
tr(E_{\alpha}H_{i})).
\end{equation}
\section{Field Equations for the Axions and the Dilatons}\label{section2}
Now that we have obtained the lagrangian \eqref{s15} explicitly in
terms of the axions and the dilatons we can derive the field
equations. We should first observe that
\begin{equation}\label{s18}
\omega=\omega(\chi_{m})\quad ,\quad \mathbf{\Omega}=
\mathbf{\Omega}(\chi_{m}).
\end{equation}
Thus we see that while the variation of \eqref{s15} with respect
to the dilatons $\{\phi^{i}\}$ is a straightforward task we should
examine the variation of $\mathbf{\Omega}$ with respect to the
axions $\{\chi^{m}\}$ from a closer look. When we vary the
lagrangian \eqref{s15} with respect to the dilaton $\phi^{k}$ we
obtain the dilatonic field equations as
\begin{equation}\label{s19}
\begin{aligned}
(-1)^{(D-1)}&d(\frac{1}{2}(\mathcal{A}_{ik}+\mathcal{A}_{ki})\ast
d\phi ^{i}+\mathcal{B}_{k\alpha}e^{\frac{1}{2}\alpha _{i}\phi
^{i}}\mathbf{\Omega}^{\alpha}_{\beta}\ast
d\chi^{\beta})\\
\\
=&\frac{1}{2}\mathcal{B}_{i\alpha}\alpha_{k}\ast d\phi ^{i}\wedge
e^{\frac{1}{2}\alpha _{i}\phi
^{i}}\mathbf{\Omega}^{\alpha}_{\beta}d\chi^{\beta}\\
\\
&+\mathcal{C}_{\alpha\beta}(\alpha_{k}+\beta_{k})e^{\frac{1}{2}\alpha
_{i}\phi ^{i}}\mathbf{\Omega}^{\alpha}_{\tau}\ast
d\chi^{\tau}\wedge e^{\frac{1}{2}\beta _{i}\phi
^{i}}\mathbf{\Omega}^{\beta}_{\gamma}d\chi^{\gamma}.
\end{aligned}
\end{equation}
Before writing down the axion field equations we will mention
about the variation of $\mathbf{\Omega}$. Firstly from \eqref{s13}
we have
\begin{equation}\label{s20}
 \omega^{\prime}\equiv\frac{\partial\omega}{\partial\chi^{m}}=K_{m},
\end{equation}
where the components of the matrix $K^{m}$ are defined as
\begin{equation}\label{s21}
(K_{m})^{\gamma}_{\beta}=K^{\gamma}_{m\beta}.
\end{equation}
Before going further we should define the adjoint representation
of $g$. The set of endomorphisms namely the linear maps on $g$
form a vector space with the addition and the scalar product
induced from $g$. They also form a Lie algebra denoted as $gl(g)$
under the product
$[\alpha,\beta]=\alpha\cdot\beta-\beta\cdot\alpha$. The
non-singular (invertible) endomorphisms of $g$ form an analytical
Lie group which we will refer as $GL(g)$. Naturally $gl(g)$ is
isomorphic to the Lie algebra of $GL(g)$. Now if we assign the map
\begin{equation}\label{s22}
ad_{X}=[X,\:\:], \quad\forall \,X \in g,
\end{equation}
such that
\begin{equation}\label{s23}
ad_{X}(Y)=[X,Y], \quad\forall \,Y \in g,
\end{equation}
then $ad_{X}$ is an endomorphism. The map
\begin{equation}\label{s24}
ad_{g}(g)\equiv ad_{X}:X\longrightarrow ad_{X},
\end{equation}
is an algebra homomorphism from $g$ into $gl(g)$ and it is called
the adjoint representation of the Lie algebra $g$. The image of
$ad_{g}(g)$ in $gl(g)$ is a subalgebra and we will denote it as
$ad(g)$. Now after introducing the elements of the adjoint
representation we can write down the partial derivative of
$e^{\omega}$ as \cite{sat,hall}
\begin{equation}\label{s25}
\begin{aligned}
\frac{\partial e^{\omega}}{\partial\chi^{m}}&=e^{\omega}(\frac{I-e^{-ad_{\omega}}}{ad_{\omega}})(\omega^{\prime})\\
\\
&=e^{\omega}(\omega^{\prime}-\frac{1}{2!}[\omega,\omega^{\prime}]+\frac{1}{3!}[\omega,[\omega,
\omega^{\prime}]]-\cdot\cdot\cdot).
\end{aligned}
\end{equation}
We should observe that the commutation series in \eqref{s25} will
terminate after a finite number of terms since from their
definitions in \eqref{s13} and \eqref{s20} both $\omega$ and
$\omega^{\prime}$ lie in the adjoint representation of
$\mathbf{n}$ which is a nilpotent Lie algebra so is its image
$ad(\mathbf{n})$ which is composed of nilpotent endomorphisms
\cite{hel}. We may see this fact as follows; if we define the
ideals
\begin{equation}\label{s26}
\varphi^{p+1}ad(\mathbf{n})=[ad(\mathbf{n}),\varphi^{p}ad(\mathbf{n})],
\end{equation}
where $\varphi^{0}ad(\mathbf{n})=ad(\mathbf{n})$ then the series
\begin{equation}\label{s27}
\varphi^{0}ad(\mathbf{n})\supset\varphi^{1}ad(\mathbf{n})\supset\varphi^{2}ad(\mathbf{n})\supset\cdot\cdot\cdot,
\end{equation}
is called the central descending series and  we observe that the
growing terms of the series \eqref{s25} belong to the growing
ideals of \eqref{s27}. Due to the nilpotency of $ad(\mathbf{n})$
\eqref{s27} terminates with $\varphi^{m}ad(\mathbf{n})=\{0\}$ for
some $m\geq $dim$(ad(\mathbf{n}))$ \cite{hel,carter,onis}. This
justifies the termination of \eqref{s25} after a finite number of
terms. The expansion of $e^{\omega}$ which is $e^{\omega}=I+\omega
+1/2! \omega^{2}+\cdot\cdot\cdot$ also terminates after a finite
number of terms since the matrix $\omega$ as an element of
$ad(\mathbf{n})$ is the representative of a nilpotent endomorphism
and for any nilpotent endomorphism $N$ $N^{k}=0$ for some $k\in
{\Bbb{Z}}$. This fact also brings termination following a finite
number of terms in the expansion of $\mathbf{\Omega}$ in
\eqref{s12}. If we vary $\mathbf{\Omega}$ with respect to the
axion $\chi^{m}$ we find that
\begin{equation}\label{s28}
\begin{aligned}
\mathcal{D}_{m}\equiv\frac{\partial\mathbf{\Omega}}{\partial\chi^{m}}\:=&\:
e^{\omega}(\frac{I-e^{-ad_{\omega}}}{ad_{\omega}})(\omega^{\prime})\omega^{-1}\\
\\
&-\mathbf{\Omega}\omega^{\prime}\omega^{-1},
\end{aligned}
\end{equation}
where we have also used
\begin{equation}\label{s29}
\frac{\partial\omega^{-1}}{\partial\chi^{m}}=-\omega^{-1}\omega^{\prime}\omega^{-1}.
\end{equation}

Now we are ready to vary the lagrangian \eqref{s15} with respect
to the axion $\chi^{m}$. Performing the variation while keeping in
mind the definitions we have introduced we obtain the axionic
field equations
\begin{equation}\label{s30}
\begin{aligned}
(-1)^{(D-1)}&d(\frac{1}{2}\mathcal{B}_{i\alpha}e^{\frac{1}{2}\alpha
_{i}\phi ^{i}}\mathbf{\Omega}^{\alpha}_{m}\ast d\phi
^{i}+\mathcal{C}_{\alpha\beta}e^{\frac{1}{2}\alpha _{i}\phi
^{i}}e^{\frac{1}{2}\beta_{i}\phi
^{i}}(\mathbf{\Omega}^{\alpha}_{\gamma}\mathbf{\Omega}^{\beta}_{m}+
\mathbf{\Omega}^{\alpha}_{m}\mathbf{\Omega}^{\beta}_{\gamma})\ast
d\chi^{\gamma})\\
\\
=&\frac{1}{2}\mathcal{B}_{i\alpha}\mathcal{D}^{\alpha}_{m\beta}e^{\frac{1}{2}\alpha
_{i}\phi ^{i}}\ast d\phi ^{i}\wedge d\chi^{\beta}\\
\\
&+\mathcal{C}_{\alpha\beta}e^{\frac{1}{2}\alpha _{i}\phi
^{i}}e^{\frac{1}{2}\beta _{i}\phi
^{i}}(\mathcal{D}^{\alpha}_{m\tau}\mathbf{\Omega}^{\beta}_{\gamma}+
\mathbf{\Omega}^{\alpha}_{\tau}\mathcal{D}^{\beta}_{m\gamma})\ast
d\chi^{\tau}\wedge d\chi^{\gamma}.
\end{aligned}
\end{equation}
\section{Conclusion}
By using the exact form of the Cartan-form in the symmetric space
sigma model lagrangian we have expressed the lagrangian explicitly
in terms of the dilatons and the axions which parameterize the
coset manifold of the SSSM in the solvable Lie algebra gauge. In
this formulation we have kept the coefficients of a generic trace
convention. Then we have directly varied this basic form of the
lagrangian to obtain the dilatonic and the axionic field
equations.

Our formulation generalizes the one in \cite{ker1,ker2} which is
based on a special trace convention. In \cite{ker1,ker2} the
lagrangian is not derived exactly, however as we have mentioned
before a dualisation method is used to find the first-order field
equations for the undetermined field strengths of the axions which
take part in the Cartan-Maurer form. Since the Cartan-Maurer form
is derived in \cite{nej1,nej2} by using the results of
\cite{nej1,nej2} in this work we exactly construct the lagrangian
and obtain the field equations directly for the coset scalars
namely the dilatons and the axions. The formulation presented in
this work is purely in algebraic terms. Our derivation expresses
both the lagrangian and the field equations in terms of the
unspecified structure constants of a generic global symmetry group
$G$ without assigning a representation. Thus the results are
powerful in applying to any specific symmetric space sigma model
example. As we mentioned in the introduction due to the special
role of the SSSM in the low energy effective string theory the
construction presented here also serves as a direct and an exact
method of calculation in the non-perturbative string dynamics.

We are also working on a similar formulation for the vielbein
formalism of the symmetric space sigma model whose construction
differs from the one presented here. Different coset
parametrizations can further be studied. Finally starting from the
field equations obtained here one can work on the first-order
formulation of the theory.

\end{document}